\documentclass[aps,pre,a4paper,twocolumn,nofootinbib,superscriptaddress]{revtex4-2}

\usepackage{graphicx}
\usepackage{hyperref}
\usepackage{latexsym}
\usepackage{amsmath, amssymb, amsthm}
\usepackage{mathtools}
\usepackage{xcolor}
\usepackage{comment}
\usepackage[shortlabels]{enumitem}

\usepackage{microtype}
\usepackage{epigraph}
\usepackage{lipsum}

\usepackage[normalem]{ulem}

\usepackage{cleveref}

\newcommand{\change}[1]{{\color{black}{#1}}}

 \newcommand{\km}{k_{\text{min}}}
 \newcommand{\be}{\begin{equation}}
 \newcommand{\ee}{\end{equation}}

\usepackage{tikz}
  \newlength\squareheight
  \DeclareRobustCommand\fourloop{\protect\tikz{\draw (0,0) rectangle (\squareheight,\squareheight)}}

  \DeclareRobustCommand\squareslash{\protect\tikz{\draw (0,0) rectangle (\squareheight,\squareheight);\draw(0,0) -- (\squareheight,\squareheight)}}
  \DeclareMathOperator\squarediv{\mkern 1mu \squareslash}

  \DeclareRobustCommand\fourclique{\protect\tikz{\draw (0,0) rectangle (\squareheight,\squareheight);\draw(0,0) -- (\squareheight,\squareheight); \draw(0,\squareheight) -- (\squareheight,0)}}

\DeclareRobustCommand{\stirlingone}[2]{\genfrac{[}{]}{0pt}{}{#1}{#2}}
\DeclareRobustCommand{\stirlingtwo}[2]{\genfrac{\{}{\}}{0pt}{}{#1}{#2}}

\newcommand{\ER}{{Erd\H{o}s-R\'{e}nyi} }

\begin{document}


\title{Strongly clustered random graphs via triadic closure: An exactly solvable model}



\author{Lorenzo Cirigliano}
\affiliation{Dipartimento di Fisica Universit\`a ``Sapienza”, P.le
  A. Moro, 2, I-00185 Rome, Italy.}

\affiliation{Centro Ricerche Enrico Fermi, Piazza del Viminale, 1,
  I-00184 Rome, Italy}

\author{Claudio Castellano}

\affiliation{Istituto dei Sistemi Complessi (ISC-CNR), Via dei Taurini
  19, I-00185 Rome, Italy}

\affiliation{Centro Ricerche Enrico Fermi, Piazza del Viminale, 1,
  I-00184 Rome, Italy}

\author{Gareth J. Baxter}
\affiliation{Departamento de F\'\i sica da Universidade de Aveiro \& I3N, Campus Universit\'ario de Santiago, 3810-193 Aveiro, Portugal}

\author{G\'abor Tim\'ar}
\affiliation{Departamento de F\'\i sica da Universidade de Aveiro \& I3N, Campus Universit\'ario de Santiago, 3810-193 Aveiro, Portugal}

\date{\today}

\begin{abstract}
  Triadic closure, the formation of a connection between two nodes in a network
  sharing a common neighbor, is considered a fundamental mechanism determining
  the clustered nature of many real-world topologies.
  In this work we define a static triadic closure (STC) model for clustered
  networks, whereby starting from an arbitrary fixed backbone network, each triad is
  closed independently with a given probability.
  Assuming a locally treelike backbone
  we derive exact expressions for the expected number
  of various small, loopy motifs (triangles, $4$-loops, diamonds
  and $4$-cliques) as a function of moments of the backbone degree distribution.
  In this way we determine how transitivity and its suitably defined
  generalizations for higher-order motifs depend on the heterogeneity of the
  original network, revealing the existence of transitions due to the interplay
  between topologically inequivalent triads in the network.
  Furthermore, under reasonable assumptions for the moments of the backbone
  network, we establish approximate relationships between motif densities,
  which we test in a large dataset of real-world networks. We find a good
  agreement, indicating that STC is a realistic mechanism
  for the generation of clustered networks, while remaining simple enough
  to be amenable to analytical treatment.
\end{abstract}


\maketitle


\section{Introduction}
\label{sec:intro}

A network representation is a powerful tool for studying a huge
variety of complex systems. Random network models, such as the
Erd\H os-R\'enyi model \cite{erdds1959random,
  erdHos1960evolution}, the more general configuration model
\cite{bollobas1980probabilistic, molloy1995critical, newman2001random}
and its various extensions \cite{dorogovtsev2022nature} have enjoyed
considerable popularity due to their amenability to mathematical
analysis. These random networks have a locally treelike structure in the
infinite size limit, which facilitates the study of branching
processes (e.g., percolation, epidemic spreading) and interacting
systems (e.g., the Ising model) on top of these substrates
\cite{dorogovtsev2008critical}.

An important feature of many real-world networks is a non-vanishing
density of short loops, in particular triangles \cite{milo2002network},
 which is at significant odds with the locally
treelike structure assumption. The propensity of node triads
to form triangles is often quantified by the local clustering
coefficient or the global transitivity~\cite{dorogovtsev2022nature}.
While the presence of clustering in networks has significant effects on processes such as percolation and epidemics \cite{serrano2006percolation,serrano2006clusteringI,serrano2006clusteringII}, both the mean local clustering coefficient and the transitivity tend to
zero in the infinite size limit of locally treelike random networks.
To account for non-vanishing clustering, i.e., a
non-vanishing density of triangles in the infinite size limit, Strauss
\cite{strauss1986general} proposed an exponential random graph model
with soft constraints on the number of edges and the number of
triangles in the network. One may introduce further constraints to
achieve specific network structures
\cite{snijders2006new, robins2007introduction,robins2007recent,battiston2020networks}.
Although such models are easily generalizable and rather flexible,
their use is impeded by the highly
non-trivial phase diagrams that can emerge~\cite{park2005solution},
making it difficult to fit real network structures and to study
dynamical models.

The latter problem is circumvented in a model proposed by Newman
\cite{newman2003properties} where each node belongs to a prescribed
number of partial cliques (fully connected subgraphs where edges are
removed with a certain probability) whose sizes are randomly
distributed. Since the building blocks---the partial cliques---only
overlap at nodes and not at edges, this model allows for analytical
treatment using generating functions to study percolation and related
processes.  The same approach is used in Refs.~\cite{newman2009random,
  miller2009percolation} where each node is prescribed an edge-degree
and a triangle-degree, and nodes are randomly joined together in pairs
to form edges (as in the original configuration model) and randomly
joined together in groups of three to form triangles. Using only
triangles and no higher-order cliques allows for better control over
the degree distribution to fit real network data.  In an elegant
approach Gleeson~\cite{gleeson2009bond} is able to fit an arbitrary
degree distribution and clustering spectrum $C_K$ (the mean
clustering coefficient of nodes of degree $K$) by prescribing an
appropriate joint degree and clique size distribution.  The modelling approach of
randomly connecting cliques or partial cliques may be generalized to
randomly connecting arbitrary subgraphs \cite{karrer2010random}. As
long as these subgraphs only overlap at nodes, standard generating
function techniques can be used to solve for various network
properties such as percolation.

Most real network structures cannot be accurately described by the
above random graph models due to the complex overlapping patterns of
clustered subgraphs. A very simple and reasonably realistic
mechanism---particularly in social networks---that is able to produce
high clustering and complex community structure is triadic
closure~\cite{rapoport1953spread}. The idea is that as a network evolves many
new links are created between nodes that share a common neighbour,
i.e., by closing triads. Triadic closure is widely considered
to be an essential mechanism of structure formation in social
networks~\cite{toivonen2009comparative, snijders2011statistical,
  gong2012evolution, klimek2013triadic, bianconi2014triadic,
  asikainen2020cumulative, peixoto2022disentangling,Holme2002,Davidsen2002,Bhat2014,Levens2022}.
Most existing
models of network formation involving triadic closure are dynamic in
nature, that is, the triadic closure mechanism is generally part of a
growth or rewiring process.
This often makes it difficult to obtain an
analytical description and to identify what network features may be
directly attributed to triadic closure.

Here we consider a minimal static model of triadic closure, whereby
given an existing (backbone) network, a fraction $f$ of existing triads
is, on average, closed.
\change{In  the case of a configuration model backbone} this model can be seen as a special case of a more general model recently considered
in Ref.~\cite{zhang2023generalized}.
\change{There, triadic closure was applied after creating a network with community structure and degree correlations, using a generalized configuration model.} 
For $f=1$ static triadic closure implies that all
nodes at distance $2$ in the backbone become nearest neighbors in the
new network.
Hence a process involving nearest neighbors on the new clustered network
is equivalent to the same process with an extended range of interaction
in the original backbone network~\cite{cirigliano2023extended}.

The simplicity of this model allows for a detailed analytical
description, which is lacking in most of the studies involving triadic
closure. By means of the generating function formalism we
derive exact results in the case where the original backbone is locally treelike.
In particular, network transitivity and densities of other higher-order motifs
(such as diamonds or loops of length 4) are expressed in terms of moments
of the backbone degree distribution. In this way we uncover the existence
of sharp transitions in clustering properties of the model as a function
of the heterogeneity of the backbone. The origin of these transitions is traced
to the competition between topologically distinct types of subgraphs created
by the closure process. A comparison between the model predictions and
a large database of networks confirms the plausibility of static triadic closure
as a generative mechanism for many real-world structures.

\section{Model description}
\label{sec:model}

\begin{figure}
   \centering
    \includegraphics[width=0.4\textwidth]{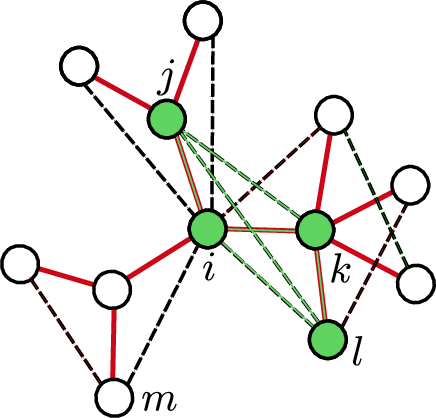}
   \caption{Pictorial representation of the STC algorithm. The edges
     of the backbone network $\mathcal{G}_0$, a tree with $N=12$ nodes
     and $E_0=11$ edges, are represented with solid red lines; the
     dashed black lines represent the edges created by the STC
     procedure. The result is a network $\mathcal{G}_f$ with $N=12$
     nodes and $E_f=21$ edges, a variety of short loops and a much more
     complex structure: many new triads, such as $(j,i,l)$
     and $(m,i,l)$, and many triangles, such as $\{i,k,l\}$, are
     created, as well as other higher-order motifs, for instance $4$-loops --
     unoriented closed paths of length $4$, e.g. $\{i,j,k,l\}$ -- and
     $4$-cliques (e.g., the one formed by the nodes shaded in green).
   }
   \label{fig:the_model}
\end{figure}

Given a graph $\mathcal{G}=(\mathcal{V},\mathcal{E})$ with
$N=|\mathcal{V}|$ nodes and $E=|\mathcal{E}|$ edges, a triad centered
on node $j$ is a sequence of three consecutive nodes
$(i,j,k)=(k,j,i)$, i.e. an unoriented path of length two made by the
edges $(i,j), (j,k) \in \mathcal{E}$. A triangle $\{i,j,k\}$ is a
closed undirected path of length three. Note that for each triangle
there are three distinct closed triads.
We define the static triadic closure (STC) mechanism as a random
process in which each triad becomes a triangle with probability $f$
through the addition of an edge joining its end nodes.
Using this STC mechanism, it is possible to build a graph in the
following way. We start from a network
$\mathcal{G}_0=(\mathcal{V},\mathcal{E}_0)$ and we write down all the
triads in it. For each triad $(i,j,k)$, we update the edge set
$\mathcal{E}_0$ by adding to it the edge $(i,k)$ with probability
$f$. When the triadic closure has been attempted on all the triads in
$\mathcal{G}_0$, the process ends. The result is a new edge list
$\mathcal{E}_f$ with $E_f=|\mathcal{E}_f|$ edges, from which we can
define a new network $\mathcal{G}_f=(\mathcal{V},\mathcal{E}_f)$ with
a rich variety of short loops and highly complex structure (see
FIG.~\ref{fig:the_model}). This algorithm defines an ensemble of
random networks, which we will refer to, with a slight abuse of
notation, as $\mathcal{G}_f$.


This algorithm may describe quite different
specific mechanisms for triadic closure.
One could
imagine for instance a triadic closure process in which a node is
inclined to be connected with one of its second neighbors with
probability $\psi$, but the triangle is only closed if both nodes
agree. This would correspond to a probability of triadic closure $f=\psi^2$.
Alternatively, the triad $(i,j,k)$ may be closed if at least
one among $i$ and $k$ likes the other, the resulting triadic closure
probability is $f=1-(1-\psi)^2$. In both cases, the STC process
described as above works with the prescription of using the
appropriate probability in place of $f$.

It is worth remarking that our definition of static triadic closure
is fully general: it is possible to generate, using this STC mechanism,
random clustered networks starting from any given backbone.
In the following sections however, we study the ensemble of graphs
$\mathcal{G}_f$, generated by static triadic closure starting from a
random, uncorrelated locally treelike backbone $\mathcal{G}_0$,
generated using the Uncorrelated
Configuration Model~\cite{catanzaro2005generation,latora2017complex}
with a prescribed degree distribution $p_k$.
We develop the theoretical framework to characterize these STC random
graphs, using generating functions to describe the properties of
$\mathcal{G}_f$ (see Appendix~\ref{appendix:A} for general definitions).
Exploiting the local treelikeness of the network
$\mathcal{G}_0$, it is possible to compute average quantities in
$\mathcal{G}_f$ in terms of averages with respect to $p_k$.
Note that in our model there are two distinct and
independent sources of randomness: the first is the random nature of
the backbone $\mathcal{G}_0$, and the second is the random nature of
the STC process.

We will denote with lower-case letters quantities which refer to the
backbone $\mathcal{G}_0$, and with capital letters quantities related
to the graph $\mathcal{G}_f$. In particular $k, r, g_0(z), g_1(z)$ and
$K, R, G_0(z), G_1(z)$ denote the degree, excess degree
and the corresponding probability generating functions, in the networks
$\mathcal{G}_0$ and $\mathcal{G}_f$, respectively.

\section{The generating function of the degree distribution}
\label{sec:degree}

We begin the study of the network ensemble $\mathcal{G}_f$ by
investigating the behavior of its degree distribution $P_K(f)$. For
the sake of brevity, we will omit the explicit dependence on $f$ of
$P_{K}(f)$ by writing $P_K$. Even if we are not able to compute $P_K$
explicitly, it is possible to determine its generating function $G_0(z)$
by expressing it in terms of the generating functions of the degree
and of the excess degree distributions in the network $\mathcal{G}_0$,
$g_0(z)$ and $g_1(z)$, respectively. The result we obtain relies on the
fact that the generating function of a sum of independent random
variables is the product of their generating functions.

Consider first the case $f=1$. $K$ is the random variable representing
the degree of a node in $\mathcal{G}_1$. Assume that this node has
degree $k$ in $\mathcal{G}_0$, and label with $i=1,\dots, k$ its first
neighbors. We can write $K = k + \sum_{i=1}^{k}r_i$, where $r_i$ is
the excess degree of first neighbor $i$.  Since the network $\mathcal{G}_0$
is uncorrelated, the generating function of the variable $K$,
conditioned on having degree $k$ in $\mathcal{G}_0$, is given by
$[zg_1(z)]^k$. Averaging then over the degree distribution
$p_k$ we get $G_0(z)=g_0(zg_1(z))$.
This standard argument can be
generalized to arbitrary $f$ by considering $K=k + \sum_{i=1}^{k}n_i$,
where $n_i$ are random variables, ranging from $0$ to $r_i$,
representing the number of new connections made with second
neighbors in the $i$-th branch. The variables $n_i$ are
independent random variables distributed according to binomial distributions
$B^{(i)}(n;r_i,f)={r_i \choose n}f^{n}(1-f)^{r_i-n}$. We can now
repeat the argument used for $f=1$ by simply conditioning not only on
$k$, but also on $r_1,\dots, r_k$. Hence, for fixed
$k, r_1, \dots, r_k$, we get $z^k\prod_{i=1}^k[(1-f+fz)^{r_i}]$, where
$(1-f+fz)^{r_i}$ is the generating function of the
binomial distribution $B^{(i)}(n_i,r_i;f)$. Averaging over
the excess degree distributions 
$q_{r_1},\dots,q_{r_k}$ and over $p_k$ we finally get for the
generating function of the degree distribution $P_K$ in
$\mathcal{G}_f$
\begin{equation}
    G_0(z)=g_0\big(zg_1(1-f+fz)\big).
    \label{eq:gen_P}
\end{equation}

The degree distribution $P_K$ may be computed, at least in principle,
from Eq.~\eqref{eq:gen_P} by differentiation, since $G_0(z)=\sum P_K
z^K$, see Appendix~\ref{sec:appendix_A_inversion}. Unfortunately, this
cannot be done explicitly in a closed form for a generic
$p_k$\footnote{It is possible to obtain an exact
  expression for $P_K$ if the backbone is a random regular network.}. However, in general
it is possible to obtain
asymptotic estimates of $P_K$ for large $K$. For instance, if
$\mathcal{G}_0$ has a power-law (PL) degree distribution $p_k\sim
k^{-\gamma}$, the resulting $\mathcal{G}_f$ is a PL network with
$P_K\sim K^{-\gamma'}$, with $\gamma'=\gamma-1$, that is the exponent is decreased by one
(see Appendix \ref{sec:appendix_A_degree} for details).

From the generating function in Eq.~\eqref{eq:gen_P} we can compute
every moment $\langle K^n \rangle$. In particular, the average degree
in $\mathcal{G}_f$ is given by
\begin{align}
\label{eq:average_degree}
\langle K \rangle&=G_0'(1) =\langle k \rangle +f\langle k(k-1) \rangle,
\end{align}
which simply states that, on average, each node is connected to a
fraction $f$ of its second neighbors, reflecting the basic mechanism
of static triadic closure. It is important to notice that if
$\mathcal{G}_0$ is a truly sparse graph, that is if $\langle k \rangle
= \mathcal{O}(1)$, see \cite{newman2018networks}, the STC procedure
creates a truly sparse graph $\mathcal{G}_f$ only if $\langle k^2
\rangle = \mathcal{O}(1)$. In some cases, such as PL with $\gamma<3$,
$\langle k^2 \rangle = \mathcal{O}(N^{\alpha})$ for some $0< \alpha
<1$, and hence $\langle K \rangle = \mathcal{O}(N^{\alpha})$. Models
defined on such networks with a slowly diverging mean degree may
exhibit a qualitatively different critical behavior from the truly
sparse case, see for instance \cite{cirigliano2023extended}.

It is useful to define the factorial moments $\mu_n$ by
\begin{equation}\label{eq:mu_n}
\mu_n = \langle k(k-1)\dots(k-n+1) \rangle = g_0^{(n)}(1)
\end{equation}
(see Appendix~\ref{sec:appendix_A_moments} for more details).

\section{Clustering}
\label{sec:clustering}

Armed with the generating function for the final degree distribution,
we may study the various properties of the new network. Since it is
the principal motivation for the model, we begin by studying
clustering properties, the presence of triangles in the network.

We first focus on the global clustering coefficient of the
network $\mathcal{G}_f$, also called transitivity,
which is the ratio between three times the total number of triangles
 to the number of triads in the network.
We then discuss the behavior of the mean local clustering coefficient
which is the ratio between the number of triangles connected to and the
number of triads centered on a particular node, averaged over all nodes.

\subsection{Transitivity}

The global clustering coefficient
is defined as~\cite{latora2017complex}
\begin{equation}
 T = \frac{3 N_{\triangle}}{N_{\wedge}},
 \label{eq:def_transitivity}
\end{equation}
where $N_{\triangle}$ and $N_{\wedge}$ denote the average total number
of triangles and triads, respectively\footnote{The factor $3$ takes
into account the fact that in each triangle there are three distinct
triads.}.

\subsubsection{General results}
Exploiting the local treelikeness of the underlying backbone
network, it is possible to exactly compute the transitivity of the
network $\mathcal{G}_f$.

The average number of triads in $\mathcal{G}_f$ can be evaluated
easily. Take a node of degree $K$ in $\mathcal{G}_f$. To form a triad,
we can pick one among $K$ of its neighbors, and then one among $K-1$
remaining other neighbors: hence such a node is the center of
$K(K-1)/2$ different triads. Averaging over the degree distribution
$P_K$ we get
\begin{align}
\nonumber
    N_{\wedge}&=N \left\langle {K \choose 2}\right\rangle =\frac{N}{2}G_0''(1)\\ \nonumber
    &=\frac{N}{2}\left[g_0''(1)(1+fg'_1(1))^2+2fg''_0(1)+f^2g_0'''(1)\right]\\
    &=\frac{N}{2}\left[\mu_2 (1+f\mu_2/\mu_1)^2+2f\mu_2+f^2 \mu_3\right],
    \label{eq:triads}
\end{align}
where we express the average over $P_K$ in terms of the derivatives
of the generating function $G_0(z)$ and we use Eq.~\eqref{eq:gen_P}.

\begin{figure}
   \centering
   \includegraphics[width=0.23\textwidth]{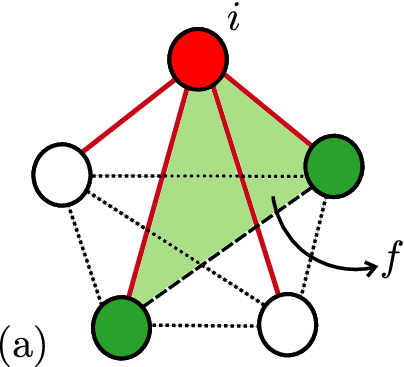}
   \includegraphics[width=0.23\textwidth]{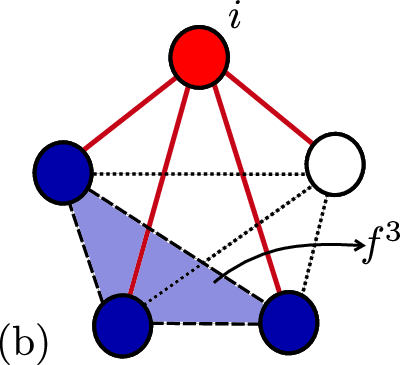}
   \caption{A visualization of the triangle counting. Consider node
     $i$ of degree $k=4$, coloured in red, and its neighbors. The
     edges in $\mathcal{G}_0$ are represented with solid red lines,
     while the black dotted lines are the edges that may be created by
     the STC mechanism and hence may appear in $\mathcal{G}_f$. We can
     distinguish between triangles of type A -- shaded in green in (a)
     -- made of two already existing edges and one new edge, and
     triangles of type B -- shaded blue in (b) -- made of three new
     edges. The total number of potential A-triangles is
     ${4 \choose 2}=6$, i.e. the number of edges in the $4$-clique composed of
     $i$'s neighbors, and each one of them is created with probability
     $f$. Over many STC realizations, there are $6f$ of them on
     average. The total number of potential B-triangles is
     ${4 \choose 3} =4$, i.e., the number of triangles in the $4$-clique composed
     of $i$'s neighbors. Each one of them appears with probability
     $f^3$, hence on average there are $4f^3$ of them. The average
     number of triangles for a node of degree $k=4$ is then $6f+4f^3$.
     Repeating this argument for all nodes we obtain
     Eq.~\eqref{eq:triangles}. Note that in this way each triangle is
     counted exactly once.}
   \label{fig:triangle}
\end{figure}

\begin{figure}[h!]
\centering
\includegraphics[width=0.48\textwidth,angle=0.]{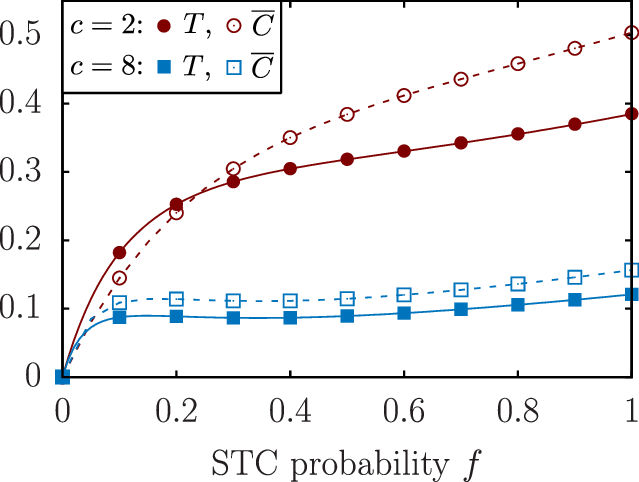}
\caption{Transitivity $T$ (filled symbols) and mean local clustering
  coefficient $\overline{C}$ (empty symbols) as a function of $f$ in
  networks $\mathcal{G}_f$ generated from ER backbones with mean
  degree $c=2$ (circles), $c=8$ (squares) and size $N=10^6$, averaged
  over $10$ realizations of the STC procedure. The continuous lines
  correspond to the exact expression for the transitivity in
  Eq.~\eqref{eq:trans_ER} and they are in perfect agreement with
  simulation results. Dashed lines are \change{not an analytic solution but simply} a guide to the
  eye.}
\label{fig:ER_transitivity}
\end{figure}

To compute the average total number of triangles we can proceed in the
following way. Consider a node $i$ of degree $k$ in $\mathcal{G}_0$,
and consider its neighborhood. We evaluate the average number of
triangles that are created in its neighborhood, and then average over
the degree distribution $p_k$. With the help of
FIG.~\ref{fig:triangle}, it is easy to see that triangles in the
neighborhood of node $i$ can be of only two types: either they are made
by joining two neighbors of node $i$ via the triadic
closure process, with probability $f$ (type A); or they are made by
joining together three neighbors of node $i$, with probability $f^3$ (type
B). Since both these types are defined with reference to the node $i$,
one may verify that the triangles will only be counted once. The
average number of triangles of type A is $fk(k-1)/2$, while the
average number of triangles of type B is $f^3k(k-1)(k-2)/3!$, hence
summing these two contributions together and averaging over $p_k$ we
obtain
\begin{align}
 N_{\triangle} &= 
\frac{Nf}{6} \left[3\mu_2+f^2\mu_3 \right].
\label{eq:triangles}
\end{align}
Substituting into Eq.~\eqref{eq:def_transitivity} we get
\begin{align}
 T &= \frac{f(3 \mu_2 +f^2 \mu_3 )}{\mu_2 (1+f\mu_2/\mu_1)^2+2f\mu_2+f^2 \mu_3}.
 \label{eq:transitivity}
\end{align}
For \ER (ER) backbones, since $\mu_n = c^n$,
Eq.~\eqref{eq:transitivity} reduces to
\begin{equation}
 T = \frac{3f+f^3c}{1+2f(1+c)+f^2c(1+c)}.
 \label{eq:trans_ER}
\end{equation}
In FIG.~\ref{fig:ER_transitivity} we compare Eq.~\eqref{eq:trans_ER}
with numerical simulations of $\mathcal{G}_f$ networks created from an
ER backbone, for different values of the original mean degree
$c$ and of the closing probability $f$. It is worth noting the
non-monotonic behavior of $T(f)$ for large, fixed $c$: for $ c>c^* =
5.7531...$, $T(f)$ admits a maximum and a minimum in the interval
$[0,1]$. This implies that from the same backbone with $c>c^*$, we can
create two graphs $\mathcal{G}_{f_1}$ and $\mathcal{G}_{f_2}$ having
the same transitivity and $f_1 \neq f_2$.

\subsubsection{The case of power-law backbones}
\label{sec:transitivity_PL}

\begin{figure}[h!]
\centering
\includegraphics[width=0.48\textwidth,angle=0.]{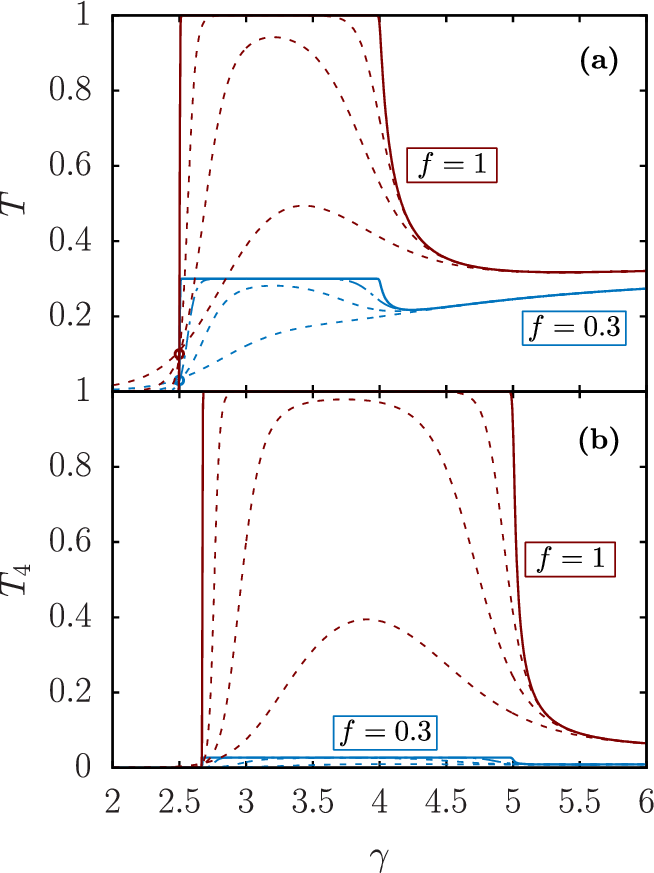}
  \caption{Analytical expressions for (a) transitivity, Eq.~\eqref{eq:transitivity}, and (b) $4$-transitivity, Eq.~\eqref{eq:generalized_transitivity} for $n=4$, in STC networks with a PL backbone, as a function of $\gamma$, for two different values of the STC probability $f$. These curves are obtained evaluating
  Eqs.~\eqref{eq:transitivity},~\eqref{eq:generalized_transitivity} within a continuous degree
  approximation, with $\km=3$. Continuous lines
  correspond to the infinite-size limit $k_c \to \infty$, dashed lines are
  for $k_c=10^3, 10^5, 10^{10}$. Circles in panel (a) correspond to
  the values for $\gamma=5/2$ given in Eq.~\eqref{eq:transitivity_asymptotic}.}
\label{fig:PL_transitivity}
\end{figure}

The case of $\mathcal{G}_f$ generated from power-law (PL) degree
distributed backbones $\mathcal{G}_0$ with $p_k \sim k^{-\gamma}$ is
of particular interest. We assume that $k\in[\km, k_c(N)]$, and that
the maximum degree grows as a power of $N$ whose value depends on the
exponent $\gamma$: $k_c(N)\sim N^{\omega(\gamma)} \to \infty$ as $N
\to \infty$ \cite{catanzaro2005generation}, with $0<\omega \le 1/2$ for $2<\gamma \le 3$
and $\omega=1/(\gamma-1)$ for $\gamma>3$.
Some moments of a
power-law distribution diverge in the infinite-size limit, depending
on the value of $\gamma$ (see Appendix~\ref{appendix:B}).
This implies that, for some
values of $\gamma$, $N_{\triangle}$ and $N_{\wedge}$ may grow faster
than linearly in $N$ \footnote{Of course, $3N_{\triangle} \leq N_{\wedge}$ always.}.
A careful analysis of
Eq.~\eqref{eq:transitivity} can be carried out, using the fact that,
when the factorial moments diverge, they are dominated by the leading term,
hence $\mu_n \sim \langle k ^n \rangle$, so that they can be used to stand in for
the moments of the degree distribution. Expressions for them are
given in Appendix~\ref{appendix:B}. For $\gamma>4$ none of the terms appearing in
Eq.~\eqref{eq:transitivity} diverge in the limit of infinite network
size, and the transitivity is a nonlinear function of $f$ and $\gamma$
which can be computed using the expressions for $\mu_n$ in
Eqs.\eqref{eq:moments1}-\eqref{eq:moments4}. More interesting is the
case $\gamma \le 4$, in which some of the terms in
Eq.~\eqref{eq:transitivity} diverge. We get, in the limit $k_c \to
\infty$,
\begin{equation}
 T \simeq \begin{cases}
          0 &\text{ for } 2<\gamma<5/2,\\
          \frac{f}{1+c(\km)} &\text{ for } \gamma=5/2,\\
          f &\text{ for }5/2<\gamma \le 4,\\
        \end{cases}
        \label{eq:transitivity_asymptotic}
\end{equation}
where $c(\km)$ is a constant, depending only on $\km$, given by the
ratio of the diverging moments $\mu_2^3/(\mu_3 \mu_1^2)$ evaluated at
$\gamma=5/2$. Within the continuous degree approximation (see Appendix
\ref{appendix:B}) $c(\km)=3\km$.
Eq.~\eqref{eq:transitivity_asymptotic} reveals a discontinuity in $T$
at $\gamma=5/2$, as observed in
FIG.~\ref{fig:PL_transitivity}(a). This discontinuous behavior occurs
because, while the number of triangles is asymptotically $\langle k^3
\rangle$, the number of triads is asymptotically proportional to
$\langle k^2 \rangle ^3 + a\langle k^3 \rangle$, where $a$ is a
constant. At $\gamma=5/2$ the dominant term in the number of triads
changes, causing the abrupt transition observed in $T$.

\begin{figure}
\centering
\includegraphics[width=0.48\textwidth,angle=0.]{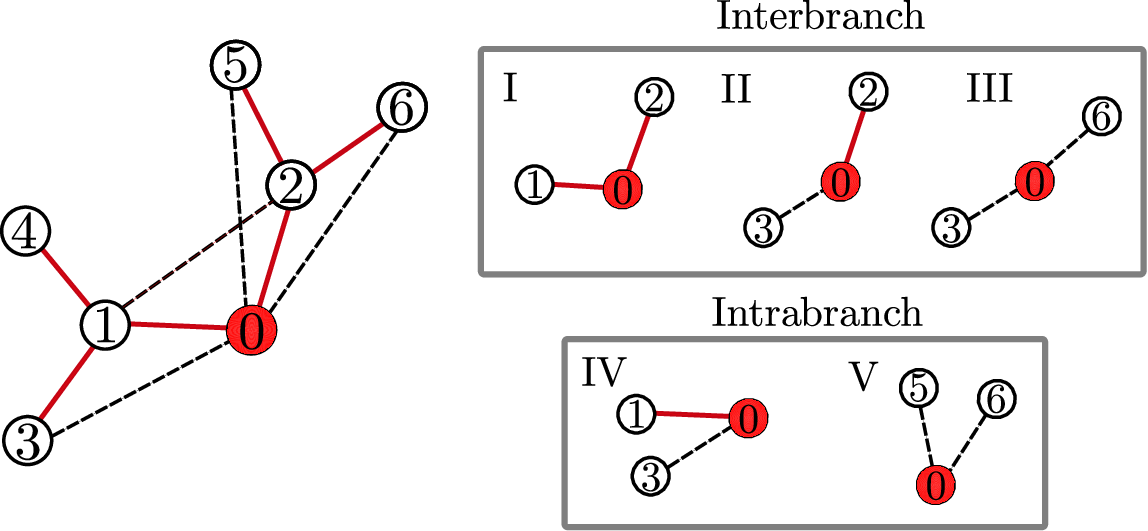}
\caption{The five classes of topologically different triads. On the
  left, a pictorial visualization of an STC process, solid red lines
  represent the edges in $\mathcal{G}_0$ and dashed black lines
  represent the edges created by the STC procedure. On the right, five
  topologically different \change{intrabranch} and \change{interbranch} triads
  centered on node $0$.}
\label{fig:triads}
\end{figure}

To understand this dual behavior of the number of triads, we identify
five classes of topologically different triads in
$\mathcal{G}_f$. Denoting with $i_0$ the center node of the triad
$(i_1,i_0,i_2)$, with the help of FIG.~\ref{fig:triads} we can
classify triads as follows.
\begin{enumerate}[label=\Roman*.]
  \item Triads in which both $i_1$ and $i_2$ were $i_0$'s neighbors in
    $\mathcal{G}_0$, such as $(1,0,2)$ in FIG.~\ref{fig:triads}. By
    construction, a fraction $f$ of these triads is closed on average.
  \item Triads in which $i_1$ and $i_2$ were $i_0$'s first and second
    neighbors, respectively, but $i_1$ and $i_2$ were not neighbors in
    $\mathcal{G}_0$, e.g. $(2,0,3)$ in FIG.~\ref{fig:triads}. These
    triads are always open in $\mathcal{G}_f$.
  \item Triads in which both $i_1$ and $i_2$ were $i_0$'s second
    neighbors in $\mathcal{G}_0$, but $i_1$ and $i_2$ did not have a
    common neighbor, such as $(3,0,6)$ in FIG.~\ref{fig:triads}. These
    triads are always open in $\mathcal{G}_f$.
  \item Triads in which $i_1$ was $i_0$'s neighbor and $i_2$ was
    $i_1$'s neighbor in $\mathcal{G}_0$, e.g. $(1,0,3)$ in
    FIG.~\ref{fig:triads}. Note that these triads are always closed.
  \item Triads in which both $i_1$ and $i_2$ were $i_0$'s second
    neighbors in $\mathcal{G}_0$, and $i_1$ and $i_2$ had a common
    neighbor, such as $(5,0,6)$ in FIG.~\ref{fig:triads}. A fraction
    $f$ of these triads is closed on average.
\end{enumerate}
We refer to triads in classes (I,II,III) as \textit{\change{interbranch}
  triads}, since they all involve nodes within different branches in
$\mathcal{G}_0$. Conversely, triads in classes (IV,V) involve nodes
within the same branch in $\mathcal{G}_0$, therefore we call them
\textit{\change{intrabranch} triads}.
It is worth noting that terms
corresponding explicitly to these five cases appear in
Eq.~\eqref{eq:triads} after expanding the term $(1+f\mu_2/\mu_1)^2$.
For $2<\gamma \le 3$, the leading contributions are both of order
$\mathcal{O}(f^2)$ and come from triads in classes III and V.
Indeed, triads from class III contribute to the denominator of $T$ with the term
\[\frac{N_{\wedge}^{\text{(III)}}}{N} = \frac{\mu_2}{2}(f\mu_2/\mu_1)^2 \sim \langle k^2 \rangle ^3 \sim k_c^{3(3-\gamma)},\]
where the first $\mu_2$ is the average number of ways of choosing two
distinct branches emanating from a random node $i$, and the factor
$(\mu_2/\mu_1)^2$ is the average number of $i$'s second neighbors in
each of such branches.
Triads from class V contribute to the denominator of $T$ with
\[\frac{N_{\wedge}^{\text{(V)}}}{N} = f^2\frac{\mu_3}{2} = \frac{f^2}{2}\mu_1(\mu_3/\mu_1) \sim \langle k^3 \rangle \sim k_c^{4-\gamma} ,\]
where the factor $\mu_1$ is the average number of branches emanating from
a random node $i$, and $\mu_3/\mu_1$ is the average number of pairs of
$i$'s second neighbors along each branch.
Note that while triads in
class V are closed with probability $f$ -- indeed they also appear in
the numerator of Eq.~\eqref{eq:transitivity} -- triads in class III
are always open. Hence if $N_{\wedge}^{(\text{III})} \ll
N_{\wedge}^{(\text{V})}$, that is for $\gamma > 5/2$, the dominant
term $N_{\wedge}^{(\text{V})}$ appears both in the numerator and the
denominator of Eq.~\eqref{eq:transitivity}, yielding $T = f$. For
$\gamma <5/2$ instead, the dominant contribution comes from open
triads $N_{\wedge}^{(\text{III})}$, and this gives $T = 0$. Then the
abrupt change in $T$ occurs when $N_{\wedge}^{(\text{III})}$ scales as
$N_{\wedge}^{(\text{V})} $, that is at $\gamma=5/2$. For
$3 < \gamma \le 4$, $N_{\wedge}^{(\text{III})}$ is finite and the dominant
contribution coming from $N_{\wedge}^{(\text{V})}$ appears both at the
numerator and at the denominator of Eq.~\eqref{eq:transitivity}, yielding
$T = f$. In other words, while for $5/2 < \gamma \le 4$ the STC mechanism
creates (almost) only closed \change{intrabranch} triads, for $\gamma <5/2$ it produces
infinitely many more open \change{interbranch} triads than closed triads.

In finite systems, one observes a slow convergence to the asymptotic results
as the system size increases, as illustrated in FIG.~\ref{fig:PL_transitivity}.

\subsection{Local clustering coefficient}

\begin{figure}
\centering
\includegraphics[width=0.48\textwidth,angle=0.]{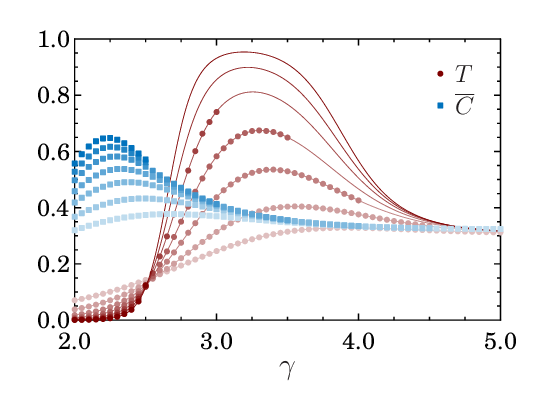}
\caption{Transitivity $T$ and mean local clustering coefficient ${\overline C}$
  as a function of $\gamma$ for PL backbones, for different values of $k_c$, with $f=1$ and $k_{\textrm{min}}=3$.
  Values of $k_c$: $100, 300, 1000, 3000, 10000, 30000, 100000$.
  Darker colors correspond to bigger values of $k_c$. The solid red lines are the exact
  transitivity values, obtained by numerically evaluating Eq. (\ref{eq:transitivity}).
  Markers correspond to simulation results. In order to correctly sample a PL
  degree distribution, for a given $k_c$, the number of nodes in the sample must fulfil $N > k_c^{\gamma-1}$.
  This criterion makes simulations involving high $k_c$ and $\gamma$ values computationally infeasible.
  For $\gamma \to \infty$ the PL backbone network converges to a random regular network of degree $k_{\textrm{min}}$.
  In this limit, since all degrees are the same, both $T$ and $\overline{C}$ are easily evaluated when $f=1$:
  $\lim_{\gamma \to \infty} T = \lim_{\gamma \to \infty} \overline{C} = 1 / k_{\textrm{min}}$. This limit is already
  clearly observed for $\gamma \approx 5$.}
\label{fig:localclustering}
\end{figure}

The local clustering coefficient $C_i=n_i^{\triangle}/n_{i}^{\wedge}$
is the ratio between the number of triangles connected to and
the number of triads centered on node $i$.  The mean
local clustering ${\overline C} = 1/N \sum_i C_i$ is another global
measure of the network structure.  While in general $T$ and $\overline
C$ do not coincide, they both tend to zero with the system size in
locally treelike random network models \cite{bianconi2005loops,van2018triadic}.
We measure numerically ${\overline C}$ in STC networks obtained from
ER backbones, and we compare it with transitivity in
FIG.~\ref{fig:ER_transitivity}. Despite some quantitative difference,
the two quantities exhibit the same qualitative behavior in this
case. A different scenario occurs in STC networks obtained from PL
backbones. We measure numerically ${\overline C}$ for various values
of $\gamma$ and report the results in FIG.~\ref{fig:localclustering}.
It turns out that the mean clustering coefficient does not exhibit the
transitions occurring for transitivity in the large-$N$ limit. Instead
$\overline{C}$ changes smoothly with $\gamma$ and also displays much
smaller finite size effects. 
We can get
some physical intuition of this qualitative difference between
$\overline{C}$ and $T$ by expressing $T$ as a weighted average of the
local clustering coefficient $C_i$\cite{latora2017complex}
\begin{equation}
 T=\frac{3\times \frac{1}{3}\sum_{i}n_i^{\triangle}}{\sum_{i}n_i^{\wedge}}=\frac{\sum_{i}n_i^{\wedge}C_i}{\sum_{i}n_i^{\wedge}}.
 \label{eq:T_weighted_average}
\end{equation}
For PL backbones with $\gamma<4$ both the numerator
and the denominator in Eq.~\eqref{eq:T_weighted_average} diverge, as we
already observed in Eq.~\eqref{eq:transitivity_asymptotic}. For
$\gamma < 5/2$, however, the divergence of the numerator is tamed
by the factor $C_i$, and the result is a vanishing transitivity.

\section{Higher-order motifs}
\label{sec:higher_order_motifs}

The complex structure of $\mathcal{G}_f$ is not limited to a large
number of triangles. Higher-order motifs, such as overlapping
triangles, loops and cliques, are also naturally created by the STC
mechanism. In this Section, by exploiting the treelike structure of
$\mathcal{G}_0$ we derive exact expressions for the number
of some higher-order motifs in $\mathcal{G}_f$.

\subsection{Diamonds and $4$-loops}
A ubiquitous feature of real social networks is not only their high
clustering, but that triangles tend to overlap, significantly altering
the dynamics in various types of processes occurring on these
networks. The frequency of triangle overlaps both in the STC model and in
real-world networks may be measured in various ways.
Here we focus on diamonds (two triangles that share a link,
that we call the ``center'' of the motif) and $4$-loops
(loops of length 4).

\subsubsection{Average number of diamonds}
\begin{figure}
   \centering
   \includegraphics[width=0.485\textwidth]{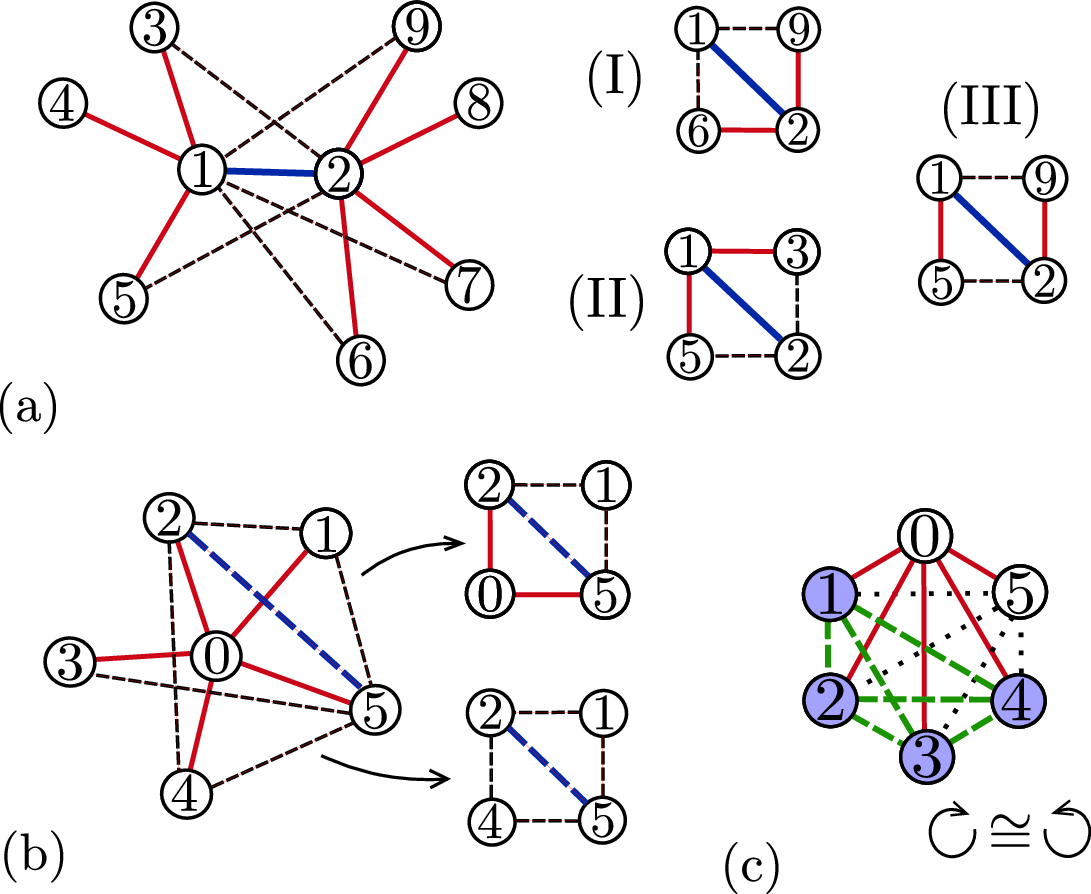}
   \caption{A pictorial visualization of diamond and $4$-loop
     counting. Solid lines correspond to links in $\mathcal{G}_0$,
     dashed lines are the links created by the STC mechanism. (a)
     diamonds centered on the old link $(1,2)$ (thicker line in blue)
     for a specific realization of the STC process. On the right, we
     identify the three types of topologically different motifs. Here
     there are $3\cdot 2/2=3$ motifs of type (I), $2/2=1$ motif of
     type (II) and $3\cdot2=6$ motifs of type (III). Summing these
     three terms and averaging gives Eq.~\eqref{eq:TT_old_impl}. Note
     that these motifs always correspond to $4$-loops, and no other
     $4$-loops with $2$ old links on the perimeter are possible, since
     $\mathcal{G}_0$ is locally treelike. (b) diamonds centered on new
     links for a particular realization of the STC process. We
     consider the neighborhood of node $0$ of degree $k_0=5$. Arrows
     indicate two topologically different diamonds centered on the new
     links $(2,5)$ (thicker line in blue). (c) Counting the $4$-loops
     with no old links on the perimeter. Here we consider the
     neighborhood of node $0$ with degree $k_ 0=5$, and the $4$-loop
     created among nodes $1,2,3,4$ (shaded in blue). There are $4$
     distinct ways of choosing the starting node, and a factor $1/2$
     arises by the simmetry under inversion. Hence taking into account
     these simmetries we have $4\cdot 3 \cdot 2/(4\cdot 2)=3$ distinct
     $4$-loops (dashed green lines): $\{1,2,3,4\}$, $\{1,3,2,4\}$,
     $\{1,3,4,2\}$. Considering that each $4$-loops has a probability
     $f^4$ of being created, averaging over the whole network we get
     Eq.~\eqref{eq:loops_0}.}
   \label{fig:TT_loops}
\end{figure}

The easiest way to compute the expected number of diamonds in
$\mathcal{G}_f$, denoted by $N_{\squareslash}$, is to write
\[N_{\squareslash}=N_{\squareslash}^{(\text{old})}+ N_{\squareslash}^{(\text{new})},\]
where $N_{\squareslash}^{(\text{old})}$ and
$N_{\squareslash}^{(\text{new})}$ are the number of diamonds centered
on old links, i.e. links that are in $\mathcal{E}_0$, and on new links,
i.e. links that are in $\mathcal{E}_f \setminus \mathcal{E}_0$,
respectively. Consider an old edge (as $(1,2)$ in FIG.~\ref{fig:TT_loops})
with end nodes $i$ and $j$.
Denoting by $n_{i \to \partial j}$ the number of new links created on average
by the STC mechanism between node $i$ and $j's$ neighbors,
we can distinguish three topologically different diamonds
centered on the old link $(i,j)$ (see FIG.~\ref{fig:TT_loops}(a)):
\begin{enumerate}[label=\Roman*.]
 \item the ones in which we take two among $j$'s neighbors which have
   become also $i$'s neighbors. There are $n_{i \to \partial j}(n_{i
     \to \partial j}-1)/2$ of them;
 \item the ones in which we take two among $i$'s neighbors which have
   become also $j$'s neighbors. There are $n_{j \to \partial i}(n_{j
     \to \partial i}-1)/2$ of them;
 \item the ones in which we consider one neighbor of node $i$ and one
   neighbor of node $j$. There are $n_{i \to \partial j}n_{j \to
     \partial i}$ of them.
\end{enumerate}
Summing these contributions and averaging over $p_k$
we get
\begin{align}
 N_{\squareslash}^{(\text{old})}
 &=\frac{N\langle k \rangle}{4}\langle \eta_{ij}(\eta_{ij}-1)\rangle
 \label{eq:TT_old_impl}
\end{align}
where $\eta_{ij}=n_{i\to \partial j}+n_{j \to \partial i}$. Using the
fact that $n_{i \to \partial j}$ and $n_{j \to \partial i}$ are
i.i.d. random variables whose probability generating function is given
by $g_1(1-f+fz)$ (see Sec.~\ref{sec:degree}), it follows that the
generating function of the probability distribution of the variables
$\eta_{ij}$ is $H(z)=\left[g_1(1-f+fz)\right]^2$. Hence we have
\begin{align}
 N_{\squareslash}^{(\text{old})}&=\frac{N\langle k \rangle}{4}H''(1)=\frac{N f^2 }{2} \left[ \mu_3+\mu_2^2/\mu_1 \right].\label{eq:TT_old}
\end{align}
Now consider a node of degree $k$ in $\mathcal{G}_0$, and a new edge
$(a,b)$ created among its neighbors. A diamond centered on this edge
can be formed only because of the creation of new links among the
other $(k-2)$ neighbors of the node, since nodes $a$ and $b$ are
second neighbors in the original network. There are on average
$fk(k-1)/2$ new links such as $(a,b)$. This link can be the center
either of a diamond with $2$ old links and $2$ new links on the
perimeter, or can be the center of a diamond with $4$ new links on the
perimeter (see FIG.~\ref{fig:TT_loops}(b)). First we consider the
diamonds with two old links on the perimeter. Fixed the link $(a,b)$,
there are $k-2$ ways of choosing the third node, and hence we have
$f^3k(k-1)(k-2)/2$ distinct motifs. In the other case, we can pick the
two other nodes to complete the diamond by choosing among the
remaining $(k-2)(k-3)$ nodes, and we can do this in $(k-2)(k-3)/2$
distinct ways. The total number of such motifs is $f^5
k(k-1)(k-2)(k-3)/4$. Averaging over $p_k$ we get

\begin{align}
    N^{(\text{new})}_{\squareslash}&= N \left [f^3\frac{\mu_3}{2}+ f^5 \frac{\mu_4}{4} \right].
\end{align}
Summing up we finally have
\begin{align}
    N_{\squareslash}=\frac{N f^2}{4} \left[2\mu_2^2/\mu_1+2(1+f)\mu_3+f^3 \mu_4 \right].
    \label{eq:TT}
\end{align}

\subsubsection{Average number of $4$-loops}
To compute the average number of $4$-loops, a simple observation is
crucial: any $4-$loop must contain either zero or two old links, by
construction. The average number of $4$-loops with two old edges,
denoted by $N^{(2)}_{\fourloop}$, is the same as the number of diamonds
centered on old links given in Eq.~\eqref{eq:TT_old}, see
FIG.~\ref{fig:TT_loops}.
The number of $4$-loops with zero old links,
denoted $N^{(0)}_{\fourloop}$, is instead simply given by (see
FIG.~\ref{fig:TT_loops}(c))
\begin{align}
    N^{(0)}_{\fourloop} = \frac{N f^4 \mu_4}{8},
    \label{eq:loops_0}
\end{align}
since we have $k(k-1)(k-2)(k-3)$ ways of picking $4$ nodes among the
neighbors of a random node of original degree $k$, and in a loop the
order in which we choose the nodes matters: we have $4$ possible ways
of starting the loop, and $2$ possible choices of orientation, which
gives the factor $1/8$. Putting these two contributions together we get
\begin{align}
    N_{\fourloop} = \frac{N f^2}{8}\left[4\mu_3 +4\mu_2^2/\mu_1 +  f^2\mu_4 \right].
    \label{eq:4_loops}
\end{align}

\subsubsection{$4$-loops made of overlapping triangles}
It is useful to define the quantity\footnote{Only for $f>0$, since for
  $f=0$ both the numerator and the denominator are zero.}
\begin{equation}
 R_{\squarediv} = \frac{N_{\squareslash}}{2N_{\fourloop}},   \label{eq:R_def}
\end{equation}
which provides a measure of how likely $4$-loops are to be made of
overlapping triangles (diamonds) in $\mathcal{G}_f$. In an arbitrary network, any diamond corresponds to
exactly one $4$-loop. Also, any $4$-loop corresponds to at most two
diamonds. Therefore, for any network $R_{\squarediv} \in [0,1]$.
From Eqs.~\eqref{eq:TT},\eqref{eq:4_loops} we get
\begin{equation}
 R_{\squarediv} = \frac{\left[2\mu_2^2/\mu_1+2(1+f)\mu_3+f^3 \mu_4 \right]}{\left[4\mu_3 +4\mu_2^2/\mu_1 +  f^2\mu_4 \right]}.
\end{equation}

From this expression, it follows that for PL networks
$R_{\squarediv} \to f$ as $k_c \to \infty$ for $2<\gamma \le 5$,
since the most divergent term $\mu_4$ is the same in the numerator and
the denominator, while for $\gamma>5$, the ratio $R_{\squarediv}$
is, in the same limit, a nontrivial function of $f$ always different
from 0 and 1. In this case (in fact, for any backbone where the first four moments
of the degree distribution are finite) we have
$\lim_{f \to 0}R_{\squarediv}=1/2$.
This means that for small but
finite $f$, each $4$-loop corresponds to exactly one diamond.

\subsection{Cliques, stars and generalized transitivity}
\label{sec:generalized_transitivity}
In this section, we present a generalization 
 of the
transitivity to higher-order motifs. It is defined by
\begin{equation}
 T_{n}=\frac{n N_{\mathbb{K}_n}}{N_{S_{n-1}}},
 \label{eq:def_generalized_transitivity}
\end{equation}
where $N_{\mathbb{K}_n}$ and $N_{S_{n-1}}$ denote the number of
$n$-cliques -- complete subgraphs with $n$ nodes, $\mathbb{K}_n$ --
and the number of $(n-1)$-stars -- subgraphs made by one node and
$n-1$ leaves, $S_{n-1}$ -- in $\mathcal{G}_f$, respectively. The
multiplicative factor $n$ takes into account the fact that for each
complete subgraph $\mathbb{K}_n$ there are $n$ distinct stars
$S_{n-1}$. Note that for $n=2$
Eq.~\eqref{eq:def_generalized_transitivity} trivially reduces to
$T_2=1$, while for $n=3$ we recover the standard transitivity $T_3=T$
as in Eq.~\eqref{eq:def_transitivity}. It is straightforward to derive
expressions for $N_{\mathbb{K}_n}$ and $N_{S_{n-1}}$ for arbitrary $n$,
following the same line of argument which led us to
Eqs.~\eqref{eq:triads},\eqref{eq:triangles}. We get
\begin{align}
    N_{S_{n-1}}&=N \left\langle {K \choose n-1}\right \rangle = \frac{N}{(n-1)!}G_0^{(n-1)}(1),
    \label{eq:n_stars}
\end{align}
where $G_0^{(n-1)}(1)$ can be computed using
Eq.~\eqref{eq:moments_with_faadibruno}, and (it is sufficient to
generalize FIG.~\ref{fig:triangle} for arbitrary $n$-cliques)
\begin{align}
\nonumber
 N_{\mathbb{K}_n} &= N\left[f^{\frac{(n-1)(n-2)}{2}} \left\langle {k \choose n-1} \right\rangle + f^{\frac{n(n-1)}{2}} \left\langle {k \choose n} \right\rangle \right]\\
&=\frac{Nf^{(n-1)(n-2)/2}}{n!} \left[n\mu_{n-1}+f^{n-1}\mu_n \right].
\label{eq:n_cliques}
\end{align}
Substituting into Eq.~\eqref{eq:def_generalized_transitivity} we get
\begin{align}
 T_{n} &= \frac{f^{(n-1)(n-2)/2}(n\mu_{n-1}+f^{n-1}\mu_n )}{M_{n-1}},
 \label{eq:generalized_transitivity}
\end{align}
where we set $M_n=\langle K(K-1)\dots(K-n+1)\rangle$. The study of
$T_{n}$ for PL backbones reveals an interesting feature. It is
possible to show that, for $k_c \gg \km$ (see Appendix
\ref{sec:appendix_C})
\begin{equation}
 M_{n-1} \sim \begin{cases}
          k_c^{n-\gamma+(n-1)(3-\gamma)}&\text{ for }2<\gamma<\gamma^*_{(n)},\\
          k_c^{n+1-\gamma} &\text{ for }\gamma^*_{(n)}<\gamma<n+1,\\
          1 &\text{ for }n+1<\gamma,\\
        \end{cases}
        \label{eq:stars_asymptotic}
\end{equation}
where
\begin{equation}
 \gamma^*_{(n)}=3-\frac{1}{n-1}.
\end{equation}
Hence for
$\gamma>n+1$ none of the terms in $T_n$ diverges. For $\gamma \leq n+1$, since $N_{\mathbb{K}_n}\sim N k_c^{n+1-\gamma}$, we get
\begin{equation}
 T_{n} \simeq \begin{cases}
          0 &\text{ for } 2<\gamma<\gamma^*_{(n)},\\
          \frac{f^{(n-1)(n-2)/2}}{1+c_n(\km)} &\text{ for } \gamma=\gamma^*_{(n)},\\
          f^{(n-1)(n-2)/2} &\text{ for }\gamma^*_{(n)} < \gamma \leq n+1,\\
        \end{cases}
        \label{eq:generalized_transitivity_asymptotic}
\end{equation}
where $c_n(\km)$ is a constant depending only on $n$ and $\km$. Eq.~\eqref{eq:generalized_transitivity_asymptotic} shows that the generalized transitivity $T_n$ exhibits a
discontinuity at $\gamma^*_{(n)}$, in perfect analogy with the
behavior of $T$ discussed in Sec.~\ref{sec:transitivity_PL}. FIG.~\ref{fig:PL_transitivity}(b) illustrates the discontinuous transition of $T_n$ with $n=4$. The
competition between two kinds of topologically distinct $(n-1)$-stars
-- those created between one node and its second neighbors in a given
branch and those created between one node and the second neighbors
reached along $n-1$ different branches, with the former dominating over
the latter for $\gamma>\gamma^*_{(n)}$ -- is responsible for the
observed discontinuous behavior of $T_n$.

\section{Motifs in real-world networks}
\label{sec:real_world}

It is 
important to compare this model's predictions with corresponding quantities
measured in real networks.
The structure of the assumed original backbone network of a given
real-world network cannot be easily inferred.
Instead we can derive some approximate relations
(given some reasonable assumptions)
between measurable quantities, which are expected to hold for any network
generated via the static triadic closure process starting from a locally
treelike backbone.

Our first assumption involves only the moments of the ``final''
network observed degree distribution:
$\langle K^m \rangle \gg \langle K^{m-1} \rangle$
for $m \geq 2$.
This condition holds in the large size limit for PL networks with
exponent $\gamma' < 3$, which is where most observed
values in real networks tend to fall.
Our second assumption involves the moments of the backbone
with those of the observed network:
$\langle K^m \rangle \approx f^m \langle k^{m+1} \rangle$.
This relation holds exactly in the large size limit when
the backbone network is PL with $\gamma > 3$ (see Appendix
\ref{sec:appendix_C}).
This would correspond to $\gamma' > 2$ in the
final observed network (see Sec.~\ref{sec:degree}),
which, again, is where most real-world networks tend to be.

These assumptions are exact in the large size limit when the backbone
network is PL with $\gamma \in [3,4]$, which corresponds to
the range $\gamma' \in [2,3]$ in the final observed network. While real networks
have a very complicated structure, are finite, and are certainly not
exactly power-law degree-distributed, we believe that our assumptions
may still be expected to be reasonable in many networks with heavy-tailed degree
distribution. Note that we do not make any assumptions about the
particular shape of the backbone degree distribution, i.e.,
we do not fit any parameters, such as the degree distribution exponent
$\gamma$.

Under the above assumptions, using the results of Sections
~\ref{sec:clustering} and~\ref{sec:higher_order_motifs} we can write
the following simple approximate expressions for the densities of the
various motifs considered in Sec.~\ref{sec:higher_order_motifs},

\begin{align}
    n_{\fourloop} &\equiv \frac{N_{\fourloop}}{N} \approx \frac{T \langle K^3 \rangle}{8},  \label{eq:RW_10} \\
    n_{\squareslash} &\equiv \frac{N_{\squareslash}}{N} \approx \frac{T^2 \langle K^3 \rangle}{4},  \label{eq:RW_20} \\
    n_{\fourclique} &\equiv \frac{N_{\mathbb{K}_4}}{N} \approx \frac{T^3 \langle K^3 \rangle}{24},  \label{eq:RW_30}
\end{align}

\noindent
that is, the densities of motifs involving four nodes depend only on
the transitivity and the third moment of the degree distribution.
Using Eq.~(\ref{eq:RW_30}) we can write the following simple
approximate expression for the $4$-transitivity,

\begin{align}
    T_4 = \frac{4 N_{\mathbb{K}_4}}{N_{S_3}} \approx T^3.  \label{eq:RW_40}
\end{align}

\begin{figure}[h!]
\centering
\includegraphics[width=0.48\textwidth,angle=0.]{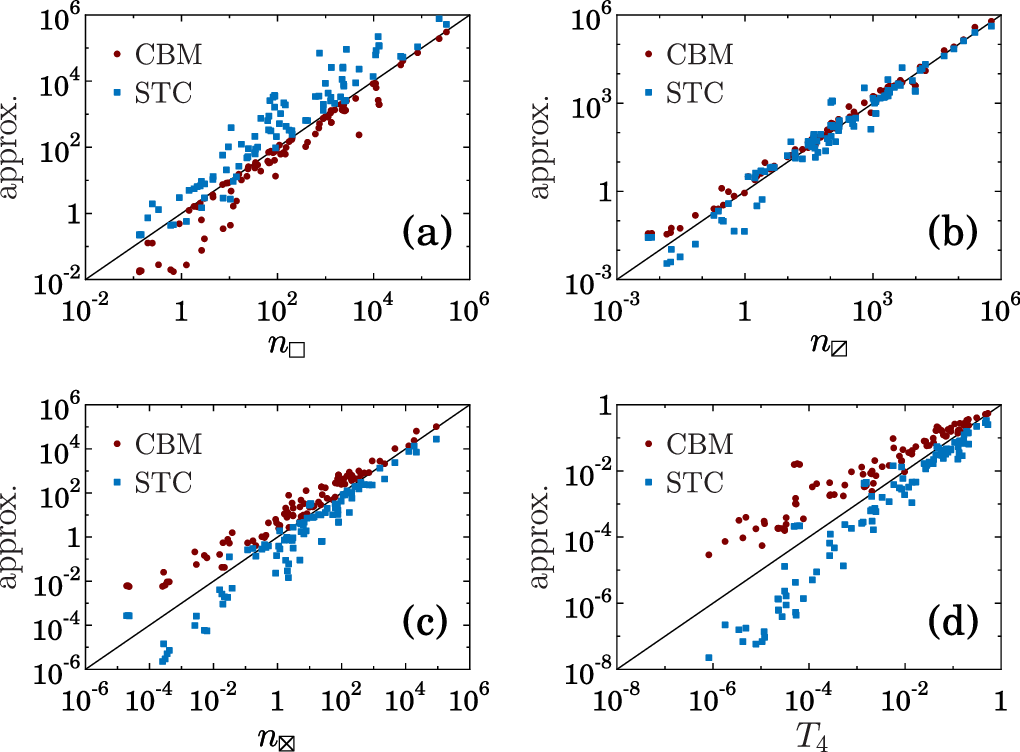}
\caption{The CBM and STC [Eqs. (\ref{eq:RW_10}-\ref{eq:RW_40})]
  approximations to the (a) density of $4$-loops, (b) density of
  diamonds, (c) density of $4$-cliques, and (d) $4$-transitivity, as
  a function of the actual observed values in a varied dataset of 95
  real-world networks. In each panel one marker corresponds to one
  network.}
\label{fig:densities}
\end{figure}

\noindent
Note that the approximate forms in Eqs.~(\ref{eq:RW_10}),
(\ref{eq:RW_20}), (\ref{eq:RW_30}) and~(\ref{eq:RW_40}) were derived
using some simple assumptions related with the moments of the degree
distributions and did not require any fitting of parameters. Thus they
constitute universal relations.  We tested these relations in a
dataset of 95 real-world networks of various nature (see
Ref.~\cite{Timar2021}~\footnote{Some of the largest networks in the
  original database could not be considered as they exceeded
   our computational resources.}
for details on the dataset).
The results are shown in FIG.~\ref{fig:densities}.  Despite the
extremely simple form of the approximate expressions, they appear to
be in reasonable agreement with empirical results for most networks.

\noindent
To further assess the validity of the approximate relations derived from
the STC model, a comparison with other models of clustered networks is
worthwhile. As mentioned in Sec.~\ref{sec:intro}, most existing
mathematically tractable models of clustered networks involve a random
linking of triangles, complete or partial cliques, or other
higher-order motifs. Unfortunately most of these models are not easily
fitted to real networks, therefore the range of applicability of
relations derived within them is not easily established. One
clique-based model where this can actually be done in a particularly
elegant manner is due to Gleeson~\cite{gleeson2009bond}. In this model
each node belongs to exactly one clique and any number of external
links. Importantly, the joint degree and clique size distribution
in this model can be exactly fitted to a given network
degree distribution and clustering spectrum $C_K$. For this reason we chose
this model, as a representative of clique-based models, to compare our
results to, and we will refer to it as~\emph{clique-based model}
(CBM). In the CBM the quantities $n_{\fourloop}$, $n_{\squareslash}$,
$n_{\fourclique}$ and $T_4$ are easily calculated as functions of the
moments of the clique-size distribution, which we fitted to the
clustering spectra
of the 95 real networks considered. The resulting
values are presented in FIG. \ref{fig:densities}.  While both the CBM
and STC models produce reasonable approximations \footnote{It is
  important to remark that the relations derived from the STC
  model are universal and did not require any parameter fitting, while
  the values for the CBM were obtained by fitting the entire degree
  distribution and clustering spectrum of a real network.}, there are
interesting differences to be considered.

The differences are due to the fact that the CBM and STC models
realize, in a sense, two opposite extreme approaches to producing
clustered networks. In the CBM, triangles only exist within
complete cliques, and this leads to an overestimation of denser
motifs, e.g., $4$-cliques and an underestimation of sparser motifs such as
$4$-loops. On the other hand, in the STC model
triangles are produced in a more homogeneous, diffuse manner,
resulting in the opposite trend: an underestimation of $4$-cliques and
an overestimation of $4$-loops{\footnote{\change{Another comparison of real world clustering statistics with Gleeson's model and a different kind of random model, using rewiring and also matching the clustering spectrum, was made in Ref. \cite{colomer2013deciphering}.}}.

A quantity that sharply highlights the difference between the two
models is the normalized ratio $R_{\squarediv}$ of the number of diamonds to
$4$-loops [see Eq.~(\ref{eq:R_def})]. This quantity always has the
trivial value $R_{\squarediv}^{\textrm{(CBM)}} = 1$ in clique-based network models (made of complete cliques),
since one $4$-loop
corresponds to exactly two diamond motifs in this case.
This property of clique-based models,
demonstrating the extreme concentration of triangles, is clearly at
odds with structures observed in real-world network data.
In the STC model, using the assumptions outlined at the beginning of
this section, this ratio is simply given by the
transitivity, $R_{\squarediv}^{\textrm{(STC)}} \approx T$.
FIG.~\ref{fig:R} shows that, for the 95 real-world networks considered,
the STC model provides reasonable approximations for $R_{\squarediv}$ in some cases,
although in general underestimates the true values.

\begin{figure}[h!]
\centering
\includegraphics[width=0.7\columnwidth,angle=0.]{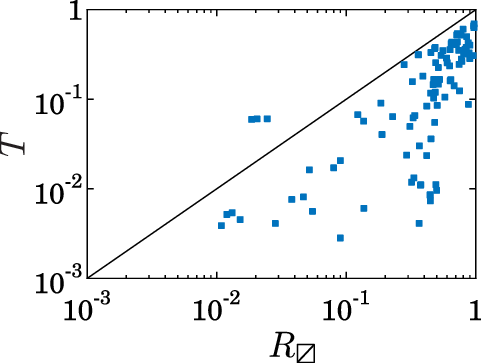}
\caption{The STC approximation ($\approx T$) to the ratio $R_{\squarediv}$ [see
    Eq. (\ref{eq:R_def})], as a function of the actual observed value
  in a varied dataset of 95 real-world networks. One
  marker corresponds to one network.}
\label{fig:R}
\end{figure}

These results suggest that more realistic versions of the STC model may be achieved
by adopting a non-homogeneous triadic closure mechanism, where the probability of
closing a triad depends on local structural properties. An obvious candidate for such
local properties to consider would be node degrees: in general one may prescribe an
arbitrary function $f(k_1,k_2,k_3)$ for the probability of closing a triad of degrees
$k_1,k_2,k_3$. This would allow for substantial control over the extent to which triangles
overlap to form loopy motifs, and would allow for the modelling of various forms of the
clustering spectrum $C_K$.

\section{Discussion and conclusions}
\label{sec:conclusions}

In this paper we studied in detail a static model of random
clustered networks based on the mechanism of triadic closure. In
particular, we start from a ``backbone'' network of arbitrary
degree distribution and, with a given probability $f$, we close
each of the existing triads.
In the case where the backbone is an uncorrelated locally treelike network,
due to its simplicity this model allows for exact analytical results
regarding clustering properties of the network.

We found an exact expression for transitivity and we showed
anomalous behavior of this quantity in large power-law degree-distributed
networks;
transitivity is equal to $0$ in the infinite size limit for degree
distribution exponent $\gamma < 5/2$ and transitivity is equal to $f$
for $5/2 < \gamma \le 4$.

This sharp transition is reminiscent of another
transition occurring for this value of $\gamma$. Indeed, the largest
eigenvalue of the adjacency matrix (spectral radius) of power-law networks
is $\langle k^2 \rangle/\langle k \rangle \sim k_c^{3-\gamma}$ for $\gamma<5/2$,
while it is $k_c^{1/2}$ for $\gamma>5/2$~\cite{Chung2003,Castellano2017}.
In that case, the transition is related to the localization
of the principal eigenvector of the adjacency matrix, either on the
K-core of maximum index or on the largest hub and its nearest
neighbors~\cite{pastor2016distinct}.
The identification of the role of \change{interbranch} and \change{intrabranch} triads
in the behavior of transitivity provides a complementary and clarifying view.
In networks with $\gamma<5/2$ a large max K-core is present.
In such structures the neighbors of nodes with large degree have many neighbors
in their turn. This is reflected by the dominance of \change{interbranch} processes in
the STC model.
For $\gamma>5/2$ instead, the spectral radius is determined by the largest hub
and its direct neighbors, which tend to have a small number of neighbors.
Correspondingly the transitivity is dominated by the formation
of connections among the hub's neighbors (\change{intrabranch} processes).

To quantify the density of higher order cliques
we defined a generalized transitivity $T_n$
as the number of $n$-cliques in the network relative to their maximum
possible number (in perfect analogy to standard transitivity).
We found an exact
expression for the general $T_n$ and showed that this quantity
undergoes a discontinuous transition---analogous to the standard
case---at $\gamma^*_{(n)} = 3-1/(n-1)$.

Using generating functions and simple combinatorial considerations we
found exact expressions for the densities of various small loopy
motifs, as functions of the first few moments of the backbone
degree distribution. Importantly, all motifs in the STC model
are produced by triangles (closed triads) overlapping in various
ways, i.e., all emerging small-scale structures are purely
induced by the random triadic closure mechanism. This circumstance
makes the STC model a useful tool to evaluate the plausibility of
the triadic closure mechanism in real-world networks. With some
reasonable assumptions about the moments of the degree distribution,
we derived some universal relations between the densities of
small loopy motifs. Specifically, we were able to
express the density of various motifs involving four nodes as a
function of transitivity and of the third moment of the degree
distribution. We showed that these remarkably simple, universal
relations, hold up reasonably well in real-world networks.

Many interesting research avenues, opened by this work, deserve further
investigation.
First, while this paper focuses on global quantities, it would be important
to understand in detail also the behavior of local quantities, such as the degree
distribution $P_K$, the local clustering coefficient $C_i$, and
degree-degree correlations.
Second, the generalization of the approach used
in~\cite{cirigliano2023extended} for percolation on the STC model with $f=1$
can provide insight into the behavior of percolation and other
processes on networks with strong clustering and many short loops.
Finally, the STC procedure can be generalized to build hypergraphs,
by considering not only the edges created by the STC mechanism but
also the triangles, as well as higher-order motifs, as hyperedges.
The motif counting analysis
developed in this work can be straightforwardly generalized in this
case, providing a nontrivial, yet exactly solvable, model for triadic
interaction.

\appendix

\section{Probability generating functions}

\subsection{General definitions}
\label{appendix:A}

Given a discrete probability distribution $f_k$, the associated
generating function is defined as
\begin{equation}
 g(z)= \langle z^k \rangle = \sum_{k} f_k z^k,
\end{equation}
where the sum is intended over the whole range of $k$ values for which
$f_k$ is defined. In case of a continuous
variable with probability density $f(k)$, we define
\begin{equation}
 g(z)=\int dk f(k)z^k.
\end{equation}

For our purposes, we consider the degree distribution $p_k = N_k/N$,
where $N_k$ is the number of nodes with degree $k$, with $k \in
[k_{\text{min}},k_c(N)]$, where $k_c(N)$ diverges in the infinite-size
limit. We denote by $g_0(z)$ the degree distribution generating
function.
Another useful distribution to consider is $q_r$, where the random
variable $r$ is the so-called excess degree, i.e., the degree of a
node at which we arrive following a randomly chosen edge excluding the
edge we arrived from.  We can express $q_r$ in terms of $p_k$. Indeed,
we can compute the probability of reaching a node of excess degree
$r$, and hence degree $r+1$, following a randomly chosen edge. This is
simply given by
\begin{equation}
 q_r = \frac{(r+1) N_{r+1}}{\sum_{r}(r+1) N_{r+1}} = \frac{(r+1) p_{r+1}}{\langle k \rangle}.
\end{equation}
This expression allows us to express the generating function
$g_1(z)=\sum_{r}q_r z^r$ in terms of $g_0$ by
\begin{equation}
 g_1(z)=\frac{g_0'(z)}{g_0'(1)}.
 \label{eq:p_q}
\end{equation}

\subsection{Analytic and asymptotic methods}
\label{sec:appendix_A_inversion}
Given a generating function $g_0(z)$ it is possible to obtain the
coefficients $p_k$, i.e. the probability distribution $p_k$, by
differentiation \cite{dorogovtsev2022nature}
\begin{equation}
 p_k=\frac{1}{k!}\frac{d^k}{dz^k}g_0(z) \big\rvert_{z=0}=\frac{1}{2\pi i} \oint_{\mathcal{C}}dz\frac{g_0(z)}{z^{k+1}},
\end{equation}
where $\mathcal{C}$ is an arbitrary path around the origin \change{in the complex $z$ plane}. It is
quite rare that this procedure can be carried out explicitly for any
$k$. Nevertheless, complex analysis developed
many tools to estimate the asymptotics of the coefficients $p_k$ for
large $k$. For instance, if $g_0(z)$ is analytic, then $g_1(z)$
is also analytic and the composition of analytic functions is analytic
too. Hence we know for sure that $P_K$ cannot exhibit a power-law tail
\cite{wilf2005generatingfunctionology}.  If instead the generating
function $g_0(z)$ exhibits a singular behavior for $z \to 1^-$, it is
possible to know the asymptotic behavior of the coefficients of the
series expansion around $z=0$, that is the asymptotics of $p_k$ for
large $k$. In particular, we recall the following result (Theorem 1
and its corollaries in \cite{flajolet1990singularity}): if
$f(z)=\sum_n f_n z^n$ is analytical in the unitary circle in the
complex plane excluding $z_0=1$, and as $z\to 1$ in this domain,
$f(z)\sim c (1-z)^{\alpha}$ for $\alpha$ real, then for noninteger
$\alpha$ we have
\begin{equation}
 f_n \sim \frac{c}{\Gamma(-\alpha)}n^{-\alpha-1}, ~~~n \to \infty.
 \label{eq:tauberian_2.0}
\end{equation}

Hence for a singular $g_0(z)$ it is sufficient to expand around 1,
using $\epsilon=1-z$ as a small parameter, to get the asymptotic form
of $p_k$.

\subsubsection{The degree distribution of $G_f$ with PL backbone}
\label{sec:appendix_A_degree}
Using Eq.~\eqref{eq:tauberian_2.0} and the expansions for the
generating functions of a PL degree distribution
\begin{align}
\nonumber
g_0(1-\varepsilon)&\simeq1-\langle k \rangle \varepsilon+\frac{1}{2}\langle k \rangle B \varepsilon^2 +C(\gamma-1)\varepsilon^{\gamma-1},\\
\nonumber
g_1(1-\varepsilon)& \simeq1-B\varepsilon +\frac{1}{2}D\varepsilon^2+C(\gamma-2)\varepsilon^{\gamma-2},
\end{align}
where $B$, $D$ and $C$ are constants depending on $\gamma$ and
$\km$\footnote{See Appendix G in \cite{cirigliano2023extended} for
  their explicit values and, in particular, their signs depending on
  the value of $\gamma$.}, from Eq.~\eqref{eq:gen_P} we get
\begin{align}
 G_0(1-\epsilon) \simeq 1 -[\langle k \rangle+f \langle k \rangle B]\epsilon+\langle k \rangle C(\gamma-2)(f\epsilon)^{\gamma-2}.
\end{align}
From Eq.~\eqref{eq:tauberian_2.0} we get for $K \to \infty$
\begin{align}
 \nonumber
 P_K &\sim \frac{\langle k \rangle C(\gamma-2)f^{\gamma-2}}{\Gamma(2-\gamma)}K^{-(\gamma-1)},
\end{align}
from which we conclude that the STC procedure on PL backbones with
exponent $\gamma$ produces (asymptotically) PL networks with
exponent $\gamma'=\gamma-1$.

\subsection{Computing averages using generating functions}
\label{sec:appendix_A_moments}
Generating functions are useful tools because they contain information
about the whole probability distribution in a very compact way
\cite{wilf2005generatingfunctionology}\footnote{From
  \cite{wilf2005generatingfunctionology},``A generating function is a
  clothesline on which we hang up a sequence of numbers
  [probabilities] for display.''}. Indeed, taking the derivatives of
the generating functions we can evaluate averages. It is easy to see
that, defining the $n$-th factorial moment
\cite{daley2003introduction} $\mu_n = \langle k(k-1)\dots(k-n+1)
\rangle$
\begin{align}
\label{eq:averages_formula_1}
 \mu_n&=  \left[\left(\frac{d}{d z}\right)^n g_0(z) \right]\biggr|_{z=1},\\
 \label{eq:averages_formula_2}
 \langle k ^n \rangle &= \left[ \left(z\frac{d}{dz}\right)^n g_0(z) \right] \biggr|_{z=1} ,
\end{align}
where we used the fact that $g_0(z)=\langle z^k \rangle =\langle e^{k
  \ln z} \rangle$ and $z d/dz = d/d(\ln z)$.

It is possible to express $\langle k^n \rangle$ as a linear
combination of $\mu_j$ for $j \leq n$ using the relation
\cite{knopf2003operator}
\begin{equation}
 \left(z\frac{d}{dz}\right)^n \!\!f(z) = \sum_{j=1}^n \stirlingtwo{n}{j}z^j \left(\frac{d}{dz}\right)^j\!\!f(z),
 \label{eq:xd/dx}
\end{equation}
where $\stirlingtwo{n}{j}$ denote the Stirling numbers of the second
kind, whose expression is given by
\cite{wilf2005generatingfunctionology, knopf2003operator}
\[\stirlingtwo{n}{j}=\sum_{i=1}^j\frac{(-1)^{j-i}i^n}{i!(j-i)!}. \]
Evaluating Eq.~\eqref{eq:xd/dx} for $f(z)=g_0(z)$ at $z=1$ we get
\begin{equation}
 \langle k^n \rangle = \sum_{j=1}^n \stirlingtwo{n}{j} \mu_j.
 \label{eq:m_vs_mu}
\end{equation}
Remarkably, Eq.~\eqref{eq:m_vs_mu} states for ER networks, for which
$\mu_j=c^j$, that
\[ \langle k^n \rangle = \sum_{j=1}^n \stirlingtwo{n}{j} c^j,\]
that is $\langle k^n \rangle$ is a power series in $c$ whose
coefficients are the Stirling numbers of the second kind.

On the other hand, it is possible to express $\mu_n$ in terms of
$\langle k^n \rangle$ for $j\leq n$ using the Stirling numbers of the
first kind $\stirlingone{n}{j}$ which are defined by the relation
\cite{knuth1997art}
\begin{equation}
 \prod_{j=0}^{n-1}(x-j)= \sum_{j=1}^n (-1)^{n-j}\stirlingone{n}{j}x^j.
 \label{eq:stirling_first}
\end{equation}
Evaluating Eq.~\eqref{eq:stirling_first} for $x=k$ and averaging over $p_k$ we get
\begin{equation}
\mu_n = \sum_{j=1}^n (-1)^{n-j}\stirlingone{n}{j} \langle k^j \rangle.
\label{eq:mu_vs_m}
\end{equation}

\section{Moments of power-law distributions}
\label{appendix:B}
Given a power law probability distribution $p_k \sim k^{-\gamma}$ with
$k \in [k_{\text{min}},k_c]$, we have
\begin{equation}
 p_k = \frac{k^{-\gamma}}{\zeta(\gamma, \km)-\zeta(\gamma, k_c)} \simeq \frac{k^{-\gamma}}{\zeta(\gamma, \km)},
\end{equation}
where $\zeta(\gamma,x)=\sum_{k\geq x} k^{-\gamma}$ is the Hurwitz zeta
function \cite{apostol1998introduction}. The moments of the
distribution are given by
\begin{equation}
 \langle k^n \rangle = \frac{\zeta(\gamma-n, \km)-\zeta(\gamma - n, k_c)}{\zeta(\gamma, \km)-\zeta(\gamma, k_c)}.
\end{equation}
To make computations less cumbersome, we can adopt the
continuous-degree approximation, in which the degree is assumed to be
a continuous variable. Note that the larger the value of $\km$, the
better this approximation works. We have
\begin{align}
 p_k&=\frac{(\gamma-1)k_{\text{min}}^{\gamma-1}}{\left[1-\left(\frac{k_{\text{min}}}{k_c}\right)^{\gamma-1}\right]}k^{-\gamma} \simeq (\gamma-1)k_{\text{min}}^{\gamma-1} k^{-\gamma},\\
  \langle k \rangle &= \frac{\gamma-1}{\gamma-2}k_{\text{min}}\left[1-\left(\frac{k_{\text{min}}}{k_c} \right)^{\gamma-2} \right] \simeq \frac{\gamma-1}{\gamma-2}k_{\text{min}},
\end{align}
for $\gamma>2$ and $k_c \gg k_{\text{min}}$, which is the case we
consider in this paper. For higher order moments of the distribution,
the result depends on the value of $\gamma$, since the $n$-th moment
may diverge. We have, for $k_c \gg \km$
\begin{align}
 \langle k^n \rangle &= \frac{(\gamma-1)k_{\text{min}}^{\gamma-1}}{\gamma-n+1}\left[k_{\text{min}}^{n-(\gamma-1)}-k_c^{n-(\gamma-1)} \right] \\
 &\underset{k_c \gg \km}\simeq
 \begin{cases}
         \frac{\gamma-1}{\gamma-1-n}k_{\text{min}}^{n}&\text{ if } n<\gamma-1,\\
         \frac{\gamma-1}{n+1-\gamma}k_{\text{min}}^{\gamma-1}k_c^{n-(\gamma-1)}&\text{ if } n>\gamma-1.
\end{cases}
\end{align}
Note that $\mu_n \sim \langle k^n \rangle$ is finite for $n<\gamma-1$:
in such a case, it is useful to have an explicit expression for
$\mu_n$, at least for $n=1,2,3,4$, those encountered in the main
text. Using Eq.~\eqref{eq:mu_vs_m} and \cite{knuth1997art} we get
\begin{align}
\label{eq:moments1}
 \mu_1 &= \langle k \rangle,\\
 \mu_2 &= \langle k^2 \rangle - \langle k \rangle ,\\
 \mu_3 &= \langle k^3 \rangle -3\langle k^2 \rangle +2\langle k \rangle ,\\
 \mu_4 &= \langle k^4 \rangle -6\langle k^3 \rangle +11\langle k^2 \rangle -6 \langle k \rangle .
\label{eq:moments4}
\end{align}
For $n>\gamma-1$ instead we have
\begin{align}
 \mu_{n} \simeq
\frac{\gamma-1}{n+1-\gamma}k_{\text{min}}^{\gamma-1}k_c^{n-(\gamma-1)}.
\end{align}

\section{High-order moments of $P_K$}
\label{sec:appendix_C}
From Eq.~\eqref{eq:gen_P} and
Eqs.~\eqref{eq:averages_formula_1}-\eqref{eq:averages_formula_2}, we
can compute, at least in principle, every average with respect to
$P_K$ in $\mathcal{G}_f$, in terms of averages with respect to
$p_k$. It is possible to obtain an explicit expression for
$M_n=\langle K(K-1)\dots (K-n+1)\rangle=G_0^{(n)}(1)$ for arbitrary
$n$ using the Fa\`a di Bruno's formula for the $n$-th derivative of a
composite function \cite{mckiernan1956derivatives}. Denoting with $D^n=(d/dz)^n$, we have
\begin{equation}
 D^n \left[f(u(z))\right]=n!\sum_{m=1}^{n}\frac{f^{(m)}(u(z))}{m!}\!\!\!\sum_{i_1+\dots+i_m=n}\prod_{j=1}^{m}\frac{u^{(i_j)}(z)}{i_j!}.
 \label{eq:faadibruno}
\end{equation}
Writing $G_0(z)=g_0(\psi(z))$, where $\psi(z)=zg_1(1-f+fz)$, it is
easy to prove by induction that
\begin{equation}
 \psi^{(n)}(z)= nf^{n-1}g_1^{(n-1)}\big(1-f+fz\big)+f^{n}zg^{(n)}_{1}\big(1-f+fz \big).
 \label{eq:psi_derivatives}
\end{equation}
From Eq.~\eqref{eq:faadibruno} with $f=g_0$ and $u=\psi$ evaluated at
$z=1$ we finally get, using
Eqs.~\eqref{eq:p_q}\eqref{eq:averages_formula_1},\eqref{eq:psi_derivatives}
\begin{equation}
 M_n=n! \! \sum_{m=1}^{n}\frac{\mu_{m}}{m!}\!\!\sum_{s_1+\dots+s_m=n}\prod_{j=1}^{m}\frac{[s_j f^{s_j-1}\mu_{s_j}+f^{s_j}\mu_{s_j+1}]}{\mu_1 s_j!}.
 \label{eq:moments_with_faadibruno}
\end{equation}
Eq.~\eqref{eq:moments_with_faadibruno} allows us to compute the
expected number of $n-1$-stars and the generalized transitivity $T_n$
for arbitrary $n>1$ (see Sec.~\ref{sec:generalized_transitivity}), but
the computation becomes soon cumbersome. Nevertheless, with
Eq.~\eqref{eq:moments_with_faadibruno} we can prove
Eq.\eqref{eq:stars_asymptotic}. In the case of PL backbones with
$2<\gamma<3$, we have $\mu_j \sim k_c^{j+1-\gamma}$ for $j>1$, while
$\mu_1 \sim 1$, and $\psi^{(j)}(1)\sim f^{j}k_c^{j+2-\gamma}$ for
$j\geq1$. From Eq.~\eqref{eq:moments_with_faadibruno} we obtain
\begin{equation}
M_n \simeq af^{n}k_c^{n+2-\gamma}+\sum_{m=2}^n b_m f^{m}k_c^{m+1-\gamma+m(3-\gamma)},
\label{eq:moments_help}
\end{equation}
where $a, b_m$ are constants. The exponent $\alpha=m+1-\gamma+m(3-\gamma)$ is a monotonically increasing
function in $m$, hence its maximum value is reached for
$m_{\alpha}=n$. Thus we have
\begin{equation}
M_n \simeq a f^{n} k_c^{n+2-\gamma} + b_n f^{n}k_c^{n+1-\gamma+n(3-\gamma)}.
\label{eq:M_n_asymptotics}
\end{equation}
The second term on the r.h.s. dominates for $2<\gamma<3-1/n$, while
the first term dominates for $\gamma>3-1/n$. For $\gamma>n+2$, none of
the terms appearing in Eq.~\eqref{eq:moments_with_faadibruno}
diverges, hence $G_0^{(n)}(1)\sim 1$. Finally, for $3-1/n < \gamma
<n+2$ the leading term is always given by $k_c^{n+2-\gamma}$: some of
the terms in the sum on the r.h.s. of are Eq.~\eqref{eq:moments_help}
$\psi^{s_j}\sim 1$, hence the exponent will be lower than $\alpha$,
and since $\alpha<n+2-\gamma$ we can conclude that the leading order
is given by the first term. With the substitution $n \to n-1$ this
yields Eq.~\eqref{eq:stars_asymptotic}.

Notice that for $3<\gamma<4$, since $M_n \sim \langle K^n \rangle $ and $ \langle k^{n+1} \rangle \sim k_c^{n+1 -\gamma}$,  Eq.~\eqref{eq:M_n_asymptotics} implies that, for $k_c \gg \km$, $\langle K^n \rangle \sim f^n \langle k^n \rangle$, as stated in Sec.~\ref{sec:real_world}.

\end{document}